\documentclass[final,3p,times]{elsarticle}
\makeatletter
\def\@makecaption#1#2{%
   \vskip 10\p@
   \setbox\@tempboxa \hbox{\bfseries{#1. #2}}%
   \ifdim \wd\@tempboxa >\hsize
       #1. #2\par
     \else
       \hbox to\hsize{\hfil\box\@tempboxa\hfil}%
   \fi}
\makeatother
\def\boxit#1{\vbox{\hbox{\kern3pt\vbox{#1\kern3pt}\kern3pt}}}
\def\figurename{\small{\bf Fig.}}

\usepackage{graphicx}
\usepackage{amssymb}

\usepackage{amsmath}
\usepackage{multirow}
\usepackage{hyperref}

\begin{document}
\begin{frontmatter}
\def\figurename{\small Fig.}


\title{Model for Constructing an Option's Portfolio with a Certain Payoff Function}





\author
{M. E. Fatyanova\corref{cor1}} \ead{fmare13@gmail.com}
\cortext[cor1]{Corresponding author}
Proceedings of 3rd International Conference on Information Technologies and Nanotechnologies - 2017, Samara, Russia, 2017: CEUR Workshop Proceedings Volume. Copyright
2017 by the author(s).

\author
{M. E. Semenov}

\address
{Tomsk Polytechnic University, 30, Lenin ave., Tomsk, 634050,
Russia}
\begin{abstract}
The portfolio optimization problem is a basic problem of financial
analysis. In the study, an optimization model for constructing an
option's portfolio with a certain payoff function has been
proposed. The model is formulated as an integer linear programming
problem and includes an objective payoff function and a system of
constraints. In order to demonstrate the performance of the
proposed model, we have constructed the portfolio on the European
call and put options of Taiwan Futures Exchange. The optimum
solution was obtained using the MATLAB software. Our approach is
quite general and has the potential to design option's portfolios
on financial markets.
\end{abstract}

\end{frontmatter}

\section{Introduction}

Interest to the options market steadily grows. In general case,
brokers are creating financial portfolios based on a combination
of standard European call and put options, cash, and the
underlying assets itself, which are not associated with an
investor's goal. Sometimes this goal is simply to insurance and
hedge \cite{Bar2012, Kaur2016}, while, in other cases, the
investor will wish to gain access to cash without currently paying
tax \cite{Hull2002} or the manager will can choice an investment
technology  \cite{Ju2013} as well as speculative purposes
\cite{Dash2007}. The option's structures appeared in 1990's and
became a popular tool of protection against falling prices
\cite{Dash2007, Garrett2013, Griffin2007}.  Most  of  the
company's hedges were conducted through  option's portfolio
(three-way collars), which involve selling a call, buying a put,
and selling a put \cite{Bar2012, Kaur2016, Griffin2007}.

In the study \cite{Bar2012} the theoretical model of zero-cost
option's strategy in hedging of sales was demonstrated. In the
model the options prices were evaluated by banks. There are seven
different cases and it is shown that the put option was not
exercised in either one of the researched cases. Therefore, it is
difficult to talk about the positive effect of hedging.

In the study \cite{Kaur2016} authors provide an empirical analysis of 
the zero cost collar option contracts for commodity  hedging  and
its financial impact analysis. Authors assessed option's portfolio
as the hedging instrument on a quantitative basis using two
scenarios in which assets prices are changed in the certain range.

In the study \cite{Ederington2002} authors proved that option's
combinations are very popular on major option markets. They show
the most popularly traded combinations in order of contract
volume: straddles, ratio spreads, vertical spreads, and strangles.
If European options were available with every single possible
strike, any \emph{smooth} payoff function could be created
\cite{Carr2002}. Authors \cite{Carr2002} gives the decomposition
formula for the replication of a certain payoff, was shown that
any twice differentiable payoff function can be written as a sum
of the payoffs from a static position on bonds, calls, and puts.
Note that authors did not impose assumption regarding the
stochastic price path. Analytical forms and graphs of the typical
payoff profiles of option trading strategies can be found in
publications \cite{Hull2002, Topaloglou2011}.

The portfolio optimization problem is a basic problem of financial
analysis. In modern portfolio theory, developed by H.~Markowitz,
investors attempt to construct the portfolios by taking some
alternatives into account: a) the portfolios with the lowest
variance correspond to their preferred expected returns and vice
versa, b) the portfolios with the highest expected returns
correspond to their preferred variance. The Markowitz optimization
is usually carried out by using historical data. The objective is
to optimize the security's weight so that the overall portfolio
variance is minimum for a given portfolio return. In this
approach, options and structured products do not have a chance to
be included into optimal mean-variance portfolios, but they have a
place in optimal behavioral portfolios \cite{Carr2002}.
Theoretical option pricing models generally assume that the
underlying asset return follows a normal distribution.

Another possible alternative is to apply some criterions to the
payoff function as well as to the initial cost of a portfolio: for
example,  market risk measure \cite{Topaloglou2011, Das2013},
probabilistic \cite{Kibzun2015}, fuzzy goal problem
\cite{Lin2016}.

In the paper \cite{Das2013} authors have proposed a method to
optimize portfolios without the normal distribution assumption of
the portfolio's return. The objective function of the portfolio
optimization problem is the expected return which is written as an
integral over product of portfolio weights underlying assets and
the joint density function of underlying asset returns. Also, the
optimization problem includes constraints on short sale as well as
the probability of not reaching thresholds.

In the study \cite{Dash2007} authors have described a class of
stock and option strategies, involving a long or short position in
a stock, combined with a long or short position in an option. It
was found that only the standard deviation, skewness, and kurtosis
of the returns distribution of the underlying stock affected the
optimal strategy, i.e. yield maximum returns. In the study
\cite{Topaloglou2011} also, first four moments (mean, variance,
skewness and kurtosis) were used to approximate the empirical
distributions of the returns. Then the authors
\cite{Topaloglou2011} have stated the multi-asset stochastic
portfolio optimization model that incorporates European options
and the portfolio has a multi-currency structure. The objective
function is to minimize the tail risk of the portfolio's value at
the end of the strategy term, $T$. In the model, the Conditional
Value-at-Risk (CVaR) metric of tail losses over the strategy term
was used. The hedge option's portfolio is optimal in the sense of
the CVaR metric. However, the proposed model depends on the
quality of a scenario tree which additionally must be test on
containing arbitrage opportunities.


Authors \cite{Topaloglou2011, Das2013} have found that optimal
behavioral portfolios are likely to include the combination of
derivative securities: put options, call options, and other
structured products. Moreover, it must be noted that portfolios
might include put options as well as call options on the same
underlying assets.

In the paper \cite{Davari-Ardakani2016} there is a proposed model
for constructing a multi-period hedged portfolio which includes an
European-type options. The objective function of optimization
problem is recorded as the difference between the expected value
of the portfolio and the expected regret of an investor. The model
takes into account, the features of long-term investment: the risk
aversion level is added into the objective function as well as the
options contract with the different time to maturity were used.

In the paper \cite{Kibzun2015} the two-step problem of optimal
investment using the probability as an optimality criterion was
studied. Various cases of distribution of returns were
investigated. It was found that the structure of the optimal
investment portfolio is almost identical despite of one or another
distribution.

In the paper \cite{Lin2016} a model for the construction of an
option's hedged portfolio was proposed under a fuzzy objective
function. The model does not explicitly takes transaction costs
into account, but the entered membership functions are aimed at
minimizing transaction lots. Thus, the authors have implicitly
tried to reduce the potential transaction costs.

In the study \cite{Ju2013} proposed a model based on using exotic
options -- lookback call options. Authors denoted that lookback
calls have positive payoffs for both maximum and minimum asset's
prices, and thus have features similar to a portfolio on call and
put options.

In the paper \cite{Moshenets2016} a computer-based system of
identifying the informed trader activities in European-style
options and their underlying asset was proposed, then the
mathematical procedure of informed trader activity monitoring was
built.


Previous studies considered the set of option's prices as a price
function of an underlying asset: $(1+\delta)\times S_t$, where
$|\delta| \le 0.1$. In contrast, in this paper the optimization is
performed over the ask- and bid-prices and their combinations. We
do not use the historical empirical distribution of returns.



Options are popular in Europe, USA, Russia, India and Taiwan
\cite{Kaur2016, Ju2013, Dash2007, Ederington2002, Das2013,
Moshenets2016}. In the paper \cite{Ederington2002} authors have
collected market data sets and shown that more than 55~percent of
the trades of 100~contracts or larger are option's combinations
and they account for almost 75~percent of the trading volume
attributable to trades of 100~contracts or larger. In the
numerical examples section of this article we will use the prices
quoted on the Taiwan Futures Exchange (TAIFEX). According to the
Report\footnote{\url{http://www.taifex.com.tw/eng/eng3/eng3_3.asp}}
in 2016 options amounted to almost 70~percent of total volume of
derivative market in Taiwan.

The purpose of this study is to construct an option's strategy
with the piecewise linear payoff function. With this in mind, the
purpose for this study can be defined as the following problem
statement: How does the personal investor's goal can be realized
with an option's portfolio? In order to answer this, two research
questions have been put forward. \begin{itemize}
    \item \textbf{Q1}: How
can the method of establishment of the option's strategy be
described from a theoretical perspective?
    \item \textbf{Q2}: How to
validate a proposed method on option's market?
\end{itemize}

The remainder of this paper is organized as follows. In Section~2
we present the main definitions and assumptions which allow us to
formulate the model as an optimization problem, including the
objective function and the constraints for option's strategy.
Section~3 then demonstrates the results of numerical experiments,
including data description, and the integer solution of the
optimization problem. Finally, Section~4 presents conclusions and
future research.

\section{Basic Definitions and Method}\label{Method}

Option contracts were originally developed and put into
circulation in order to reduce financial risks (hedging,
insurance), associated with the underlying assets. In addition to
existing standard option strategies \cite{Hull2002} actively
developing trading models, oriented to the objectives of a
particular trader (speculative trading) \cite{Dash2007},
construction of synthetic positions \cite{Topaloglou2011}, .

An \emph{option} is a contract that will give an option holder a
right to buy (or sell) the underlying asset. An \emph{options
premium} is the amount of money that investors pay for a call or
put option.

A \emph{call option} is a contract that will give its holder a
right, but not the obligation, to purchase at a specified time, in
the future, certain identified underlying assets at a previously
agreed price. A \emph{put option} is a security that will give its
holder a right, but not the obligation, to sell at a specified
time, in the future, certain identified underlying assets at a
previously agreed price.

The \emph{strike price} (exercise price) is defined as the price
at which the holder of an options can buy (in the case of a call
option) or sell (in the case of a put option) the underlying
security when the option is exercised.

An \emph{American option} may be exercised at the discretion of
the option buyer at any time before the option expires. In
contrast, a \emph{European option} can only be exercised on the
day the contract expires.

A \emph{covered option} involves the purchase of an underlying
asset (equity, bond or currency) and the writing a call option on
that same asset. Short selling is the sale of a security that is
not owned by a trader, or that a trader has borrowed.

A \emph{zero-cost option strategy} is an option trading strategy
in which one could take a free options position for hedging or
speculating in equity, forex and commodity markets.

There are main types of option's portfolio in real-world
applications in terms of the time to expiration: American-,
European-type options \cite{Hajizadeh2016}, and their combination.

The various option combinations represent strategies designed to
exploit expected changes of the options values: the price of the
underlying asset, its volatility, the time to expiration, the
risk-free interest rate, the cost to enter \cite{Hull2002,
Ederington2002}. There is a large number of possible option
combinations. When there are only two possible strike prices and
two times to expiration we can design $36$ combinations on one
call and one put which may be either bought or sold. This number
of combinations will increase significantly when any options
values (strike prices, times to expiration, underlying assets)
will be expanded insignificantly.

In this study we propose the strategy which involves European call
and put options on the same underlying asset with the same
maturity date $T$, but different strikes in a series. Let
$K_c=\{k^i_c \in \mathbb{Z}_{>0}, i \in I\}$ and $K_p=\{k^i_p \in
\mathbb{Z}_{>0}, i \in I\}$ be the call and put strikes, $K_c$,
$K_p$ are the increasing sequence of positive integers:
$$k^i_c<k^{i+1}_c, k^i_p<k^{i+1}_p, \forall i=1, 2, \ldots,
n-1,$$ $K=\{K_c \cup K_p\}$ is the set of unique strikes,
$\mathbb{Z}_{>0}=\{x \in \mathbb{Z}: x > 0 \}$ denotes the set
of positive integers. 
Let the number of call and put options be $$X_c=\{x_i^c \in
\mathbb{Z}: L\le x_i^c \le U,L<0,U>0, i \in I\},$$ $$X_p=\{x_i^p
\in \mathbb{Z}: L \le x_i^p \le U,L<0,U>0, i \in I\},$$ with
$x_i^c, x_i^p>0$ for buying, $x_i^c, x_i^p<0$ for selling, if
$x_i^c$ or $x_i^p$ equal to $0$ it means that the contract does
not include in the portfolio, $L$ and $U$ represents the lower and
upper bounds of the integer search space, respectively, $I=\{1,2,
\ldots, n\}$ is the set of indices, and $S_t$ is a price of the
underlying asset at calendar time, $0 \le t \le T$, $\hat{S}_T$ is
an expected (forecasting) price of the underlying asset at the end
of the strategy term, $T$ (single period). Prices $S_t, \hat{S}_T
\in \mathbb{R}_{>0}$, where $\mathbb{R}_{>0}=\{x \in \mathbb{R}: x
> 0 \}$ denotes the set of positive real numbers. We assume that
the initial capital $W$ is given and that no funds are added to or
extracted from the portfolio, $0 \le t \le T$. 

 In order to
determine the number of call and put options $X=\{X_c$, $X_p\}$
for the implementation of the individual investor goal we propose
the following assumptions that have impact on the payoff $V(T,X)$
and an initial cost $C(t,X)$ of portfolio:
\begin{itemize}
    \item (i) the strategy should have protection on the downside and upside
of strike prices,
    \item (ii) the strategy should effectively limit the upside earnings and
downside risk with a maximal loss, $\mathcal{L}$,
    \item (iii) the strategy should have the certain initial cost
to enter $C(t,X)$ at time $t=0$.
\end{itemize}

 %

\subsection{Objective Function of Payoff}

To establish the strategy we propose to use a combination long and
short positions in put and call contracts based on the same
underlying asset with different strike prices. We consider that
one can take a static position (buy-and-hold), and the portfolio
can include $x_i^c$, $x_i^p$ units of European call and put
options, $i \in I$. Its value at time $t$ is given by the formula
\begin{equation}\label{eqV}
V(T,X)= \sum_{i=1}^{n} x_i^c (S_t - k_c^i)^{+} + x_i^p (k_p^i -
S_t)^{+}, 
\end{equation}
the first term is the value of the call option payoff and the
second is the value of the put option payoff, and $$X^+=\max(X, 0).$$ 

Let the best ask- and bid-prices for buying and selling of call
and put options at time $t=0$ be

$$A_c=\{a_c^i \in
\mathbb{R}_{>0}, i \in I\}, B_c=\{b_c^i \in \mathbb{R}_{>0}, i\in
I\},$$ $$A_p=\{a_p^i \in \mathbb{R}_{>0}, i \in I\}, B_p=\{b_p^i
\in \mathbb{R}_{>0}, i \in I\},$$
 which we will name the \emph{input constants}:
$$b_c^i<a_c^i, b_p^i<a_p^i, i\in I \text{ and}$$
$$a_c^i > a_c^{i+1}, a_p^i < a_p^{i+1}, b_c^i > b_c^{i+1}, b_p^i < b_p^{i+1},$$
$i=1,2, \ldots, n-1$. The initial cost portfolio at time $t=0$ can
be expressed as
\begin{equation}\label{eqC}
C(t, X)=\sum_{i=1}^{n} x_i^c \cdot g_c(x_i^c)+ x_i^p \cdot
g_p(x_i^p),
\end{equation}
where the functions $g_c(x_i^c)$ and $g_p(x_i^p)$ are defined as
\begin{equation}\label{eq_g_c}
g_c(x_i^c) =
\begin{cases}
a_c^i \in A_c, &\text{if } x_i^c>0, \\
b_c^i, \in B_c, &\text{if } x_i^c\le0,\\
\end{cases}
\end{equation}
\begin{equation}\label{eq_g_p}
g_p(x_i^p) =
\begin{cases}
a_p^i \in A_p, &\text{if } x_i^p>0, \\
b_p^i, \in B_p, &\text{if } x_i^p\le0,\\
\end{cases}
\end{equation}
 $i \in I$. Thus taking into account Eq.~(\ref{eqV}) and
Eq.~(\ref{eqC}) the overall profit and loss at time $T$ will be
the final payoff minus the initial cost. So the objective function
can be expressed as
\begin{eqnarray}\label{eqF}
F(X)=V(T, X)-C(t, X)
=\sum_{i=1}^{n} x_i^c((\hat{S}_T-k_c^i)^{+} -g_c(x_i^c))
+x_i^p((k_p^i - \hat{S}_T)^{+} -g_p(x_i^p)),
\end{eqnarray}
which is a linear function of the \emph{decision} variable $X =\{X_c, X_p\}$. 
The objective functional $F(X)$ maps the entire stochastic process
(cash flow) to a single real number $$F(X): \mathbb{Z}^n \mapsto
\mathbb{R}, \text{where } \mathbb{Z}^n \mapsto \mathbb{Z}\times  \stackrel{n}{\cdots}  \times \mathbb{Z}.$$ 

\subsection{Selection of Input Parameters}

Using the conditional functions $g_c(\cdot)$ and $g_p(\cdot)$ in
the objective function Eq.~(\ref{eqF}) leads us to solve a
sequence of optimization problems. There are four input parameter
values for each call and put options: ($A_c$ or $B_c$) and ($A_p$
or $B_p$). In this case, the number of permutations based on the
selection between alternative prices (ask or bid) and possible
contracts (call or put) equal to
$N=2^n \times 2^n= 2^{2n}$. 
In the numerical examples section of this article
(Section~\ref{Example}) we will use the ask- and bid-prices quoted
on the Taiwan Futures Exchange (TAIFEX).

Let $\mathcal{C}$ denote the set of all $2 \times n$-tuples of
elements of given ordered sets of ask- and bid-prices $A_c$,
$B_c$, $A_p$ and $B_p$. The set $\mathcal{C}$ can be expressed as
\begin{align}\label{setC}
\mathcal{C}=\{(x_1, x_2,\dots, x_n, y_1, y_2, \ldots, y_n): 
x_i=a_c^i \text{ or } b_c^i, y_i =a_p^i \text{ or } b_p^i, 
i \in I\}.
\end{align}
%
%
Thus $\mathcal{C}=\{c_1, c_2, \ldots, c_{N} \}$ is the set of
ordered permutations without replacement of two $n$-elements sets
$A_c$, $B_c$ and two $n$-elements sets $A_p$, $B_p$.


The calculation of the portfolio's terminal payoff under each
price combination
in the vector notation takes the form:
\begin{align}\label{eqFvector}
\max_{X}\{X_c^{\top}((\hat{S}_T-K_c)^{+} - G_c(X_c))
+ X_p^{\top}((K_p - \hat{S}_T)^{+} - G_p(X_p))\} 
=\max_{X}\{F_{\mathcal{C}}(X)\},
\end{align}
where $\mathcal{C}$ denotes the set of ordered permutations of
model input constants Eq.~(\ref{setC}), and $G_c(X_c)$, $G_p(X_p)$
are the vector notation of conditional functions defined in
Eq.~(\ref{eq_g_c}) and  Eq.~(\ref{eq_g_p}). The objective function
Eq.~(\ref{eqFvector}) maximizes the option's payoff over the
holding period $[0, T]$.

\subsection{System of Constraints}

Each objective function $F_{\mathcal{C}}(X)$ from the
set~Eq.~(\ref{eqFvector}) is the piecewise linear function. Taking
into account the assumptions mentioned in Section~\ref{Method} we
should determine the slope of the objective function
Eq.~(\ref{eqFvector}) in the unique strike
intervals. 
We separately investigate the intervals $0\le S_T \le k_1$,
$k_2\le S_T\le k_3$, $\ldots$, $k_m\le
S_T<+\infty$, 
here $k_1=\min(K_c, K_p)$ is the smallest strike and
$k_m=\max(K_c, K_p)$
is the largest strike. 

The horizontal slope of the function (\ref{eqFvector}) in the
first closed interval $[0, k_1]$ and the left-closed interval
$[k_m, +\infty)$ are specified respectively by:
\begin{eqnarray}\label{eq1}
\sum_{i=1}^n x_i^c&=&0, \mbox{ if }  S_T \in [0, k_1], \\
\sum_{i=1}^n x_i^p&=&0, \mbox{ if } S_T \in [k_m, +\infty).
\end{eqnarray}
Positive and negative slopes of the function (\ref{eqFvector}) in
the interior intervals $[k_q, k_{q+1}]$ are provided by:
\begin{equation}\label{eq2}
\sum_{i: k_c^i \leq k_q } {x_i^c} -\sum_{j: k_p^j \geq k_{q+1} }
{x_j^p} \text{ is }
\begin{cases}
\geq 0, &\text{if $k_q \le k$}, \\
\leq 0, &\text{if $k_q > k$},
\end{cases}
\end{equation}
here $k \in K$ is an inflection point of the function
(\ref{eqFvector}).




The next balance constraint defines the bound of the downside risk
with a maximal loss, $\mathcal{L}$, over the holding period $0\le
t \le T$:
\begin{equation}\label{eq3}
V(T, X)= -\mathcal{L}.
\end{equation}

The objective function value (\ref{eqFvector}) at the terminal
time $T$ must be positive:
\begin{equation}\label{eq4}
V(T, X)>0, S_T=\hat{S}_T.
\end{equation}

The model has the liquidity constraints, we assume that an
investor can buy at least $U$ and sell at least $L$ contracts at
each
strike price $k_c^i, k_p^i \in K$: 
\begin{equation}\label{eq5}
L \le x_i^c, x_i^p \le U, L <0, U > 0, i \in I.
\end{equation}

The inflection point $k$ introduced in Eq.~(\ref{eq2}), the
maximal loss $\mathcal{L}$, 
and the expected (forecasting) price $\hat{S}_T$ Eq.~(\ref{eq4})
should be specified by an investor. Thus to address research
question \textbf{Q1} we formulated the integer linear programming
of option's portfolio selection problem~(\ref{eqFvector}), subject
to the portfolio constraints~(\ref{eq1})-(\ref{eq5}).

We assume that an investor can use the money received from the
sale of some contracts to buy other contracts in the portfolio,
then the initial cost of portfolio $C(t, X)$ Eq.~(\ref{eqC}) can
be either a positive number or zero, or even negative number. 
In the Section~\ref{Example} we will represent the series of
numerical experiments for these three cases. Thus, the system of
constraints~(\ref{eq1})-(\ref{eq5}) can be (optionally) extended
with the constraint on the initial cost of portfolio:
$$C(t, X)
\gtreqqless 0.$$
Another series of numerical experiments will be conducted to
define the sensitivity of the solution on the liquidity
constraints Eq.~(\ref{eq5}).


\begin{figure*}
  \centering
  \includegraphics[width=0.95\textwidth]{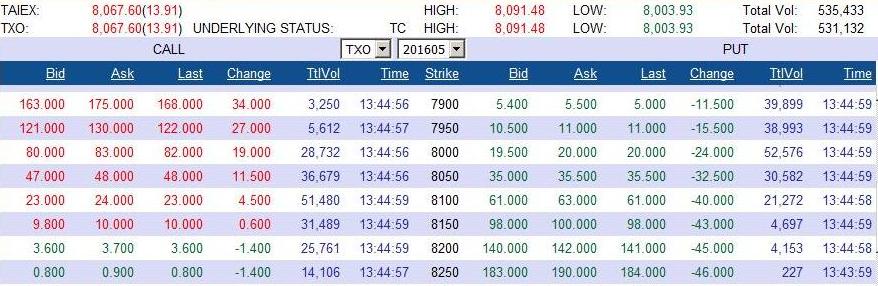}\\
  \caption{\textbf{Option Snapshot Quotes of the Taiwan Futures Exchange at
May 16, 2016.}}\label{fig1}
\end{figure*}

\section{Data Collection and Processing}\label{Example}

To address research question \textbf{Q2} we apply the optimization
problem (\ref{eqFvector})--(\ref{eq5}) and construct the option's
strategy with the certain payoff function on the derivatives of
Taiwan Futures Exchange. All options in the Taiwan's market are
European-style. The expiration periods of TXO options have spot
month, the next two months, and the next two quarterly calendar
months\footnote{\url{http://www.taifex.com.tw/eng/eng4/Calendar.asp}}.
We will be designing option's portfolio from the daily closing
ask- and bid-prices for TXO
options\footnote{\url{http://www.taifex.com.tw/eng/eng2/TXO.asp}}.
We select TXO
options; they are liquid assets and comprise above
$60\%$ of trading volume of TAIFEX. 
Here we have taken a single date, May 16, 2016, selected at
random, to illustrate the portfolio design in practice.

The strikes and the ask- and bid-prices of options are required
inputs to the portfolio optimization model. The available TXO
prices are denominated in New Taiwan Dollars (NTD). The TAIFEX
index closed at $8,067.60$ on May 16, 2016 (Fig.~\ref{fig1}), and
the May options contracts expired $9$~days later, on May 25, 2016.
The option price equals to $S_0=8,067.60$~NTD at May 16, 2016.
Next, we will consider two cases for the expected price,
$\hat{S}_T$. Suppose that the price will significant move up to 1)
$\hat{S}_T = 8{,}300.00$~NTD, 2) $\hat{S}_T = 8{,}400.00$~NTD at
May 25, 2016.

The investor then wants to monetize his position, $C(t,X)=100$~NTD
at the time of purchase, $t=0$, and to limit the maximum loss by
$\mathcal{L}=-100$~NTD, if the price of underlying asset will come
out of the certain range from $8{,}000$ to $8{,}400$~NTD,
respectively. The margin requirement for short positions and
transaction costs are not accounted.
%
%
\begin{table}[t]
\centering \caption{Optimal portfolios with the different initial
costs, $C$, and the expected price $\hat{S}_T$, NTD} \label{tab1}
\begin{tabular}{crrrrrrrrrrrr}
\hline \multicolumn{1}{l}{} & \multicolumn{6}{c}{$\hat{S}_T=8400$}
& \multicolumn{6}{c}{$\hat{S}_T=8300$} \\ \hline
\multirow{2}{*}{\begin{tabular}[c]{@{}c@{}}Strike \\
Price\end{tabular}} &
\multicolumn{2}{c}{$C=100$} &
\multicolumn{2}{c}{$C=-100$}&
\multicolumn{2}{c}{$C=0$}   &
\multicolumn{2}{c}{$C=100$} &
\multicolumn{2}{c}{$C=-100$} &
\multicolumn{2}{c}{$C=0$} \\ \cline{2-13}
 & \multicolumn{1}{c}{Call} & \multicolumn{1}{c}{Put} & \multicolumn{1}{c}{Call} & \multicolumn{1}{c}{Put} & \multicolumn{1}{c}{Call} & \multicolumn{1}{c}{Put} & \multicolumn{1}{l}{Call} & \multicolumn{1}{l}{Put} & \multicolumn{1}{l}{Call} & \multicolumn{1}{l}{Put} & \multicolumn{1}{l}{Call} & \multicolumn{1}{l}{Put} \\ \hline
7850 &  & 0 &  & 0 &  & 0 &  & 0 &  & 0 &  & 0 \\
7950 &  & 0 &  & 0 &  & 0 &  & 0 &  & 0 &  & 0 \\
\textbf{8050} & 4 & -3 & 7 & -6 & 3 & -3 & 7 & -7 & 3 & -2 & 5 & -5 \\
8150 & -8 & 8 & -10 & 9 & 1 & 2 & -8 & 9 & -2 & 4 & -4 & 6 \\
8250 & 10 & -5 & 4 & 0 & -5 & 6 & 4 & 3 & 0 & 0 & 0 & 4 \\
8350 & -8 & 0 & -3 & -3 & -5 & -5 & -3 & -5 & -1 & -2 & -5 & -5 \\
8400 & -5 &  & -2 &  & 2 &  & -9 &  & -7 &  & -2 &  \\
8500 &  7 &  &  4 &  & 4 &  &9  &  & 7 &  & 6 &  \\ \hline
\multicolumn{1}{l}{$\max\limits_{X}\{F_{\mathcal{C}}(X)\}$} &
\multicolumn{2}{c}{\textbf{700}} & \multicolumn{2}{c}{400} &
\multicolumn{2}{c}{600} & \multicolumn{2}{c}{\textbf{400}} &
\multicolumn{2}{c}{250} & \multicolumn{2}{c}{300} \\ \hline
\multicolumn{1}{l}{Total number of contracts}&
\multicolumn{2}{c}{58} & \multicolumn{2}{c}{48} &
\multicolumn{2}{c}{36} & \multicolumn{2}{c}{64} &
\multicolumn{2}{c}{28} & \multicolumn{2}{c}{42}
\\ \hline
\end{tabular}
\end{table}
\begin{table}[]
\centering \caption{\textbf{Optimal portfolios with the different
liquidity constraints, $|L|=U$, contracts, and the expected price
$\hat{S}_T$, NTD}} \label{tab2}
\begin{tabular}{crrrrrrrrrrrr}
\hline \multicolumn{1}{l}{} & \multicolumn{6}{c}{$\hat{S}_T=8400$}
& \multicolumn{6}{c}{$\hat{S}_T=8300$} \\ \hline
\multirow{2}{*}{\begin{tabular}[c]{@{}c@{}}Strike \\
Price\end{tabular}} &
\multicolumn{2}{c}{$|L|=10^{ a)}$} & \multicolumn{2}{c}{$|L|=50$}
& \multicolumn{2}{c}{$|L|=100$} & \multicolumn{2}{c}{$|L|=10^{
a)}$} & \multicolumn{2}{c}{$|L|=50$}       &
\multicolumn{2}{c}{$|L|=100$}
\\
\cline{2-13}  & \multicolumn{1}{c}{Call} & \multicolumn{1}{c}{Put}
& \multicolumn{1}{c}{Call} & \multicolumn{1}{c}{Put} &
\multicolumn{1}{c}{Call} & \multicolumn{1}{c}{Put} &
\multicolumn{1}{l}{Call} & \multicolumn{1}{l}{Put} &
\multicolumn{1}{l}{Call} & \multicolumn{1}{l}{Put} &
\multicolumn{1}{l}{Call} & \multicolumn{1}{l}{Put} \\ \hline
7850           &       & 0    &        & 0    &       & 0     & &0 & &0 & &0              \\
7950           &       & 0    &        & 0    &       & 0     & &0 & &0 & &0              \\
\textbf{8050}  & 4     & -3   & -24    & 24   & -51   & 51    & 7& -7& -24& 24& -50& 50   \\
8150           & -8    & 8    & 50     & -50  & 100   & -100  & -8& 9& 47& -47& 100& -100 \\
8250           & 10    & -5   & -11    & 30   & -4    & 49    & 4& 3& -4& 22& -12& 50     \\
8350           & -8    & 0    & -27    & -4   & -89   & 0     & -3& -5& -21& 1& -40& 0                 \\
8400           & -5    &      & -1     &      & 21    &       & -9& & -15& & -35&                   \\
8500           & 7     &      & 13     &      & 23    &       & 9&
& 17& & 37&
\\
\hline
\multicolumn{1}{l}{$\max\limits_{X}\{F_{\mathcal{C}}(X)\}$}&
\multicolumn{2}{c}{700} & \multicolumn{2}{c}{1800} &
\multicolumn{2}{c}{4400} & \multicolumn{2}{c}{400} &
\multicolumn{2}{c}{800} & \multicolumn{2}{c}{1800}
\\ \hline
\multicolumn{1}{l}{Total number of contracts}&
\multicolumn{2}{c}{58} & \multicolumn{2}{c}{234} &
\multicolumn{2}{c}{488} & \multicolumn{2}{c}{64} &
\multicolumn{2}{c}{222} & \multicolumn{2}{c}{474}
\\ \hline
\multicolumn{13}{l}{$^{a)}$  The column $|L|=10$ corresponds to
the column $C=100$ of Table~\ref{tab1}.}
\end{tabular}
\end{table}
To establish the proposed strategy we use $12$ strike prices:
sequential $n=6$ strike prices are corresponding to call
$$K_c=(\textbf{8050}, 8150, 8250, 8350, 8400, 8500)$$ and sequential $n=6$
strike prices are corresponding to put $$K_p=(7850, 7950,
\textbf{8050}, 8150, 8250, 8350)$$ at the same expiration date
May~25, 2016. The central strike of the option is $K=8050$, is
marked with bold. The number of combinations of ask- and bid-price
for the $12$ options equal to $2^n \times 2^n =2^6 \times 2^6 =
4096$, thus the cardinality of set of feasible portfolios
$|\mathcal{C}|=4096$. The cardinality of the set $|K|=|K_p \cup
K_c|$ is $8$, therefore, we have $7$ pairs of sequential strike
prices $[k_q,
k_{q+1}]$, and that these pairs produce 
the system of $7$ inequalities from Eq.~(\ref{eq2}), and we should
add two equalities on the first closed interval and the
left-closed interval from  Eq.~(\ref{eq1}). In our example, we
assumed that one can buy or sell at least $L=-10$, $U=10$
contracts at each strike price. This assumption does not limit our
approach because the total volume (\emph{TtlVol}, Fig.~\ref{fig1})
is bigger for all strike prices. Then we calculate the price for
call and put in accordance with the specific strike
(Fig.~\ref{fig1}). Next, we maximized the objective function
$F_\mathcal{C}(X)$, proposed in Eq.~(\ref{eqFvector}) with the
system of constraints~(\ref{eq1})--(\ref{eq5}). The optimal
portfolio was obtained in approximately two minutes on a personal
computer, using the MATLAB software.

As a result, the optimum solution in the case $\hat{S}_T=8400$,
$C=100$ is
$$X=(\underbrace{4,-8,10,-8,5,7}_{\text{call}},
     \underbrace{0,0,-3,8,-5,0}_{\text{put}})$$
the first six elements correspond to call options, the second six
-- to put options, the total number of contracts are 58 out of
which 34 are for buying and 24 are for selling, with objective
function value equal to 700~NTD (bold in Table~\ref{tab1}). The
optimum solution in the case $\hat{S}_T=8300$, $C=100$ is
$$X=(\underbrace{7,-8,4,-3,-9,9}_{\text{call}},
     \underbrace{0,0,-7,9,3,-5}_{\text{put}})$$
-- the total number of contracts are 64 out of which 32 are for
buying and 32 are for selling, with objective function value equal
to 400~NTD (bold in Table~\ref{tab1}). From the Table~\ref{tab1}
one can see that the maximum values of the payoff function are
achieved at $C = 100$~NTD, and the minimum values at $C =
-100$~NTD.

At the end of the strategy term, May 25, 2016, the price of
underlying index increased to $S_T=8{,}396.20$~NTD. The forecast
come true, the amount of loss was limited by the maximal loss
$\mathcal{L}=-100$.

\subsection{The Sensitivity of the Solution to the Constraints
Variation}

The initial cost $C(t, X)$ Eq.~(\ref{eqC}) can be either a
positive number or zero, or even negative number. We calculated
the alternative portfolios with the different initial costs
$C(t,X)=\{-100, 0, 100\}$ with fixed liquidity constraints
$|L|=U=10$~Eq.~(\ref{eq5}), and then the values of liquidity
constraints were varied $|L|=U=\{10, 50, 100\}$ with the fixed
initial cost $C(t, X)=100$, results are represent in
Tables~\ref{tab1},~\ref{tab2}. Table~\ref{tab2} shows the strong
dependence: with the increase in the number of liquidity
constraints, i.e. $|L|=U$, the maximum value of the payoff
function grows too.
\begin{figure}[t]
  \centering
  \includegraphics[width=1.0\textwidth]{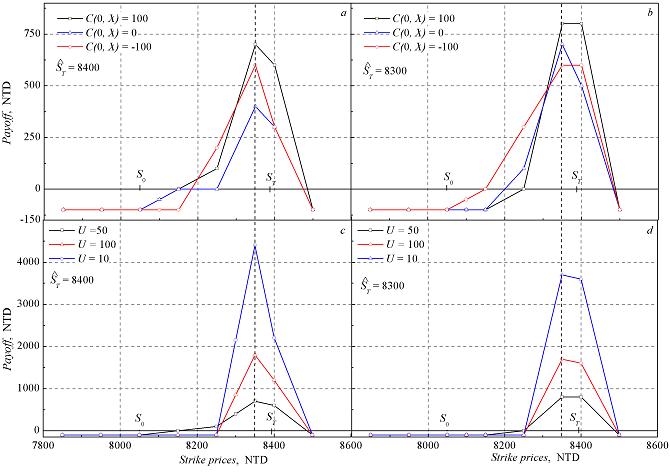}
  \caption{\textbf{Payoff functions of proposed option's
portfolios, underlying $S_0=8{,}067.60$~NTD. Different initial
costs $C(t,X): 100, 0, -100$~NTD: a) $\hat{S}_T = 8{,}400.00$~NTD,
b) $\hat{S}_T = 8{,}300.00$~NTD. Different liquidity constraints
$L$, $U$, contracts: 10, 50, 100: c) $\hat{S}_T = 8{,}400.00$~NTD,
d) $\hat{S}_T = 8{,}300.00$~NTD.}}\label{fig2}
\end{figure}
Fig.~\ref{fig2} show the payoff functions from the proposed
option's portfolio, taking into account, the different values of:
a), b)~the initial cost $C(t, X)$, Eq.~(\ref{eqC}) and c),
d)~liquidity constraints $L$, $U$, Eq.~(\ref{eq5}), respectively.
Our strike prices are the $x$-axis and the payoff functions are
the $y$-axis.

The expansion of the boundaries on the liquidity constraints
$|L|=U \in \{10,50,100\}$ makes the options portfolio more
attractive from the point of view of the terminal payoff amount.
On the other hand, there is the difficulty of forming a strategy
in view of the buy/sell of a sufficiently large number of
underlying assets, a transaction which is difficult to implement
for a short time.




\section{Conclusion and Future Research} 

In this study, the description of the method of construction of
the option's strategy with the piecewise linear payoff function is
carried out. We take into account the next set of assumptions of
the proposed option's strategy: strategy should effectively limit
the upside earnings and downside risks; strategy should have an
initial cost to enter; strategy should have protection on the
downside and upside of the underlying asset price.

To address research question \textbf{Q1} we have formulated the
mathematical model as an integer linear programming problem which
includes the system of constraints in the form of equalities and
inequalities. The optimum solution was obtained using the MATLAB
software.

To address research question \textbf{Q2} we have demonstrated the
possibility of the proposed model on the European-style TXO
options of Taiwan Futures Exchange.

In the study we do not use the historical empirical distribution
of returns. Our approach is statical, quite general and has the
potential to design option's portfolios on financial markets.

In option's strategies, in addition to the forecast of the price
of the underlying asset, various parameters can be taken into
account: exercise price, volatility, the time to expiration, the
risk-free interest rate, option premium, transaction costs. In
this case, even an insignificant change of the parameter's values
can lead to a significant change in the number of possible
combinations of option's strategies. In our numerical experiments
the total number of contracts for different cases varieties from
28 to 64 (Table~\ref{tab1}) and from 58 to 488 (Table~\ref{tab2}).
In papers \cite{Davari-Ardakani2016, Goyal2007, Primbs2009}, it is
noted that transactional costs in the dynamic management of the
portfolio of options are one of the key factors without which it
is impossible to talk about the feasibility of using the proposed
models.  The use of option strategies, including covered options,
leads to deformation of the initial distribution of returns -- it
becomes truncated and asymmetric. The payoff of the option's
portfolio is asymmetric and non-linear, therefore, from the point
of view of risk management. The use of a portfolio with various
options contracts is preferable and effective, but the problem of
choosing an optimal portfolio is significantly complicated too.

In this paper, we used the European type options in a series only.
Pricing and valuation of the American option, even the
single-asset option, is a hard problem in a quantitative finance
\cite{Hajizadeh2016, Mitchell2014}. One of possible approach in
the pricing and considering early exercise of the American option
is dynamic programming.

The further research of our study can be continued in the
following directions. At first, it is a portfolio optimization
under transaction costs (exchange commissions, brokerage
fees) 
and the margin requirement for short positions which are essential
in the options market. At second, it is using options with
different time to the expiration. At third, it is necessary to
extend the system of constraints and add the budget constraint.
Such extensions allows us to make the proposed approach more
realistic
and flexible.


\section{Acknowledgments}

Thanks to the editor and referees for several comments and
suggestions that were instrumental in improving the paper. We are
grateful to Dr. Sergey V.~Kurochkin (National Research University
Higher School of Economics, Russia) and Mr.~Ashu Prakash (Indian
Institute of Technology, Kanpur, India) for valuable comments and
suggestions that improved the work and resulted in a better
presentation of the material.






\providecommand*{\BibDash}{---}

\bibliographystyle{gost2003}
\bibliography{sample1}

\begin{thebibliography}{10}
\def\selectlanguageifdefined#1{
\expandafter\ifx\csname date#1\endcsname\relax
\else\selectlanguage{#1}\fi}
\providecommand*{\href}[2]{{\small #2}}
\providecommand*{\url}[1]{{\small #1}}
\providecommand*{\BibUrl}[1]{\url{#1}}
\providecommand{\BibAnnote}[1]{}
\providecommand*{\BibEmph}[1]{#1}
\ProvideTextCommandDefault{\cyrdash}{\iflanguage{russian}{\hbox
  to.8em{--\hss--}}{\textemdash}}
\providecommand*{\BibDash}{\ifdim\lastskip>0pt\unskip\nobreak\hskip.2em plus
  0.1em\fi
\cyrdash\hskip.2em plus 0.1em\ignorespaces}
\renewcommand{\newblock}{\ignorespaces}

\bibitem{Bar2012}
\selectlanguageifdefined{english}
\BibEmph{Bartonova,~M.} Hedging of sales by zero-cost collar and its financial
  impact [Text]~/ Marie~Bartonova~// \BibEmph{Journal of Competitiveness}.
  \BibDash
\newblock 2012. \BibDash
\newblock Vol.~4, no.~2. \BibDash
\newblock P.~111--127. \BibDash
\newblock Access mode: \BibUrl{http://www.cjournal.cz/files/99.pdf}.

\bibitem{Kaur2016}
\selectlanguageifdefined{english}
\BibEmph{Kaur,~A.} Commodity hedging through zero-cost collar and its financial
  impact [Text]~/ Amandeep~Kaur, Amandeep~Singh~Rattol~// \BibEmph{Journal of
  Energy and Management}. \BibDash
\newblock 2016. \BibDash
\newblock Vol.~1, no.~1. \BibDash
\newblock P.~44--57. \BibDash
\newblock Access mode: \BibUrl{http://www.pdpu.ac.in/downloads/4.
  commodity_hedging.pdf}.

\bibitem{Hull2002}
\selectlanguageifdefined{english}
\BibEmph{Hull,~J.} Fundamentals of Futures and Options Markets [Text]~/
  J.C.~Hull. \BibDash
\newblock New Jersey~: Financial Times, 2002.

\bibitem{Ju2013}
\selectlanguageifdefined{english}
\BibEmph{Ju,~N.} Options, option repricing in managerial compensation: Their
  effects on corporate investment risk [Text]~/ Nengjiu~Ju, Hayne~Leland,
  Lemma~W.~Senbet~// \BibEmph{Journal of Corporate Finance}. \BibDash
\newblock 2013. \BibDash
\newblock Access mode:
  \BibUrl{http://dx.doi.org/10.1016/j.jcorpfin.2013.11.003}.

\bibitem{Dash2007}
\selectlanguageifdefined{english}
A study of optimal stock and options strategies [Text]~/ M.~Dash, V.~Kavitha,
  K.M.~Deepa, S.~Sindhu~// \BibEmph{Social Science Research Network}. \BibDash
\newblock 2007. \BibDash
\newblock Access mode: \BibUrl{http://dx.doi.org/10.2139/ssrn.1293203}.

\bibitem{Garrett2013}
\selectlanguageifdefined{english}
\BibEmph{Garrett,~S.} An Introduction to the Mathematics of Finance. A
  Deterministic Approach [Text]~/ S.J.~Garrett. \BibDash
\newblock [S.\ l.]~: Elsevier Ltd., 2013.

\bibitem{Griffin2007}
\selectlanguageifdefined{english}
\BibEmph{Griffin,~B.} Review of collar options for cotton industry [Text]~/
  B.~Griffin~// \BibEmph{Chemonics International Inc.} \BibDash
\newblock 2007. \BibDash
\newblock Access mode: \BibUrl{http://pdf.usaid.gov/pdf{_}docs/PNADK976.pdf}.

\bibitem{Ederington2002}
\selectlanguageifdefined{english}
\BibEmph{Ederington,~L.} Option spread and combination trading [Text]. \BibDash
\newblock 2002. \BibDash
\newblock Access mode:
  \BibUrl{http://optionsoffice.ru/wp-content/uploads/2013/08/Option-spread-combination-trading-{\_}-Research-paper.pdf}.

\bibitem{Carr2002}
\selectlanguageifdefined{english}
\BibEmph{Carr,~P.} Towards a theory of volatility trading [Text]~/ P.~Carr,
  D.~Madan~// Volatility: New Estimation Techniques for Pricing Derivatives~/
  Ed.\ by\ R.A.~Jarrow. \BibDash
\newblock London~: Risk Books, 1998. \BibDash
\newblock P.~417--427.

\bibitem{Topaloglou2011}
\selectlanguageifdefined{english}
\BibEmph{Topaloglou,~N.} Optimizing international portfolios with options and
  forwards [Text]~/ N.~Topaloglou, H.~Vladimirou, S.A.~Zenios~//
  \BibEmph{Journal of Banking and Finance}. \BibDash
\newblock 2011. \BibDash
\newblock Vol.~35. \BibDash
\newblock P.~3188--3201.

\bibitem{Das2013}
\selectlanguageifdefined{english}
\BibEmph{Das,~S.} Options and structured products in behavioral portfolios
  [Text]~/ S.R.~Das, M.~Statman~// \BibEmph{Journal of Economic Dynamics and
  Control}. \BibDash
\newblock 2013. \BibDash
\newblock Vol.~37. \BibDash
\newblock P.~137--153.

\bibitem{Kibzun2015}
\selectlanguageifdefined{english}
\BibEmph{Kibzun,~A.} The two-step problem of investment portfolio selection
  from two risk assets via the probability criterion [Text]~/ A.I.~Kibzun,
  A.N.~Ignatov~// \BibEmph{Automation and Remote Control}. \BibDash
\newblock 2015. \BibDash
\newblock Vol.~76. \BibDash
\newblock P.~1201--1220.

\bibitem{Lin2016}
\selectlanguageifdefined{english}
\BibEmph{Lin,~C.-C.} Hedging an option portfolio with minimum transaction lots:
  A fuzzy goal programming problem [Text]~/ Chang-Chun~Lin, Yi-Ting~Liu,
  An-Pin~Chen~// \BibEmph{Applied Soft Computing}. \BibDash
\newblock 2016. \BibDash
\newblock Vol.~47. \BibDash
\newblock P.~295--303.

\bibitem{Davari-Ardakani2016}
\selectlanguageifdefined{english}
\BibEmph{Davari-Ardakani,~H.} Multistage portfolio optimization with stocks and
  options [Text]~/ Hamed~Davari-Ardakani, Majid~Aminnayeri, Abbas~Seifi~//
  \BibEmph{International Transactions in Operational Research}. \BibDash
\newblock 2016. \BibDash
\newblock Vol.~23, no.~3. \BibDash
\newblock P.~593--622.

\bibitem{Moshenets2016}
\selectlanguageifdefined{english}
\BibEmph{Moshenets,~M.~K.} Automatic system of detecting informed trading
  activities in european-style options [Text]~/ M.~K.~Moshenets,
  O.L.~Kritski~// \BibEmph{Journal of Engineering and Applied Sciences}.
  \BibDash
\newblock 2016. \BibDash
\newblock Vol.~11, no.~9. \BibDash
\newblock P.~5727--5731.

\bibitem{Hajizadeh2016}
\selectlanguageifdefined{english}
\BibEmph{Hajizadeh,~E.} Optimized radial basis function neural network for
  improving approximate dynamic programming in pricing high dimensional options
  [Text]~/ Ehsan~Hajizadeh, Masoud~Mahootchi~// \BibEmph{Neural Computing and
  Applications}. \BibDash
\newblock 2016. \BibDash
\newblock P.~1--12.

\bibitem{Goyal2007}
\selectlanguageifdefined{english}
\BibEmph{Goyal,~A.} Option returns and volatility mispricing [Text]~/
  Amit~Goyal, Alessio~Saretto~// \BibEmph{Social Science Research Network}.
  \BibDash
\newblock 2007. \BibDash
\newblock Access mode: \BibUrl{https://ssrn.com/abstract=889947}.

\bibitem{Primbs2009}
\selectlanguageifdefined{english}
\BibEmph{Primbs,~J.~A.} Dynamic hedging of basket options under proportional
  transaction costs using receding horizoncontrol [Text]~/ James~A.~Primbs~//
  \BibEmph{International Journal of Control}. \BibDash
\newblock 2009. \BibDash
\newblock Vol. 192, no.~3. \BibDash
\newblock P.~1841--1855.

\bibitem{Mitchell2014}
\selectlanguageifdefined{english}
\BibEmph{Mitchell,~D.} Boundary evolution equations for american options
  [Text]~/ D.~Mitchell, J.~Goodman, K~Muthuraman~// \BibEmph{Math. Finance}.
  \BibDash
\newblock 2014. \BibDash
\newblock Vol.~24, no.~3. \BibDash
\newblock P.~505?--532.

\end{thebibliography}



\end{document}